\documentclass[aps,twocolumn,superscriptaddress,floatfix]{revtex4-1}

\usepackage{graphicx,color}
\usepackage{braket}
\usepackage[colorlinks,urlcolor=blue,citecolor=blue]{hyperref}
\usepackage{calc}
\usepackage{amsmath}
\usepackage{amssymb}

\newcommand{\sun}{SU($N$) }
\makeatletter
\begin{document}

\title{SU(3) fermions on the honeycomb lattice at 1/3-filling}

\author{Sangwoo S.~\surname{Chung}} \affiliation{Institute for Theoretical Physics Amsterdam and Delta Institute for Theoretical Physics, University of Amsterdam, Science Park 904, 1098 XH Amsterdam, The Netherlands}

\author{Philippe~\surname{Corboz}}  \affiliation{Institute for Theoretical Physics Amsterdam and Delta Institute for Theoretical Physics, University of Amsterdam, Science Park 904, 1098 XH Amsterdam, The Netherlands}

\begin{abstract}
   SU($N$) symmetric fermions on a lattice, which can be realized in ultracold-atom-based quantum simulators, have very promising prospects for realizing exotic states of matter. Here we present the ground state phase diagram of the repulsive SU(3) Hubbard model on a honeycomb lattice at 1/3 filling obtained from  infinite projected entangled pair states tensor network calculations. In the strongly interacting limit the ground state has plaquette order. Upon decreasing the interaction strength $U/t$ we find a first order transition at $U/t=7.2(2)$ into a dimerized, color-ordered state, which extends down to $U/t=4.5(5)$ at which the Mott transition occurs and the ground state becomes uniform. Our results may serve as a prediction and benchmark for future quantum simulators of SU(3) fermions.
\end{abstract}

\maketitle

\section{Introduction}
Quantum simulators based on ultracold fermionic alkaline earth atoms in an optical lattice offer a very exciting route to realize exotic phases of \sun symmetric fermions which can be described by an \sun Hubbard model~\cite{Cazalilla2009,Gorshkov2010,Taie2012,Scazza2014,Zhang2014,Cazalilla2014,Hofrichter2016,ozawa18}.
By exploiting different nuclear spin states of the alkaline earth atoms, systems with up to $N=10$ different flavors (or colors) of fermions can be obtained. Thanks to an almost-perfect decoupling of the nuclear spin from the electronic angular momentum, the interactions are essentially independent of the nuclear spin, giving rise to a large \sun symmetry. 
Another route to realization of \sun systems (or SO(5) \cite{wu2003}) is provided by exploiting different hyperfine states of alkali atoms~\cite{wu2003, honerkamp2004,rapp2008}.

On the theory side, substantial progress has been made in recent years in predicting the ground states of these systems in the Mott insulating state in the strongly interacting limit, where the system is effectively described by an \sun Heisenberg model. On two-dimensional (2D) lattices, a rich variety of exotic ground states has been found, including various states with color order~\cite{toth2010, Bauer12, corboz11-su4},  generalized valence-bond solids~\cite{read1989,read1990,harada2003,arovas2008,hermele2011,Zhao2012,Corboz12_simplex,Corboz2013, nataf16}, algebraic $N$-flavor spin liquids~\cite{affleck1988-sun,marston1989,Assaad05,Corboz12_su4}, and chiral spin liquids~\cite{hermele2009,hermele2011,szirmaiG2011}. 

An even richer behavior can be expected in  \sun Hubbard models, where enhanced charge fluctuations in the intermediate interaction range may give rise to new interesting phases. Accurate calculations would also be desirable for a more direct comparison and prediction for future quantum simulators. However,  the Hubbard model is also substantially more challenging than the Heisenberg model, especially in the strongly correlated regime in two dimensions. While there has been progress in recent years based on dynamical mean-field theory\cite{Gorelik2009,inaba10,Inaba2012,Yanatori2016,Koga2017,DelRe2018},  variational Monte Carlo~\cite{Assaad05,Rapp2011,boos18}, Quantum Monte Carlo for exceptional cases without a sign problem~\cite{Lang2013,cai13,Wang2014}, and other approximate approaches~\cite{honerkamp2004,Hasunuma2016},  obtaining controlled and systematic predictions using unbiased numerical approaches remains a central challenge in general. In contrast, in the one-dimensional case  \sun Hubbard models have been accurately studied using matrix product states~\cite{Buchta2007,Zhao2007,Szirmai2008,Capponi2008,Molina2009,Roux2009,Ulbricht2010,Nonne2010,Manmana2011,Barcza2015,Bois2015,Barbarino2016,Szirmai2017,Weishselbaum2018}.

\begin{figure}[]  
\includegraphics[width=\columnwidth]{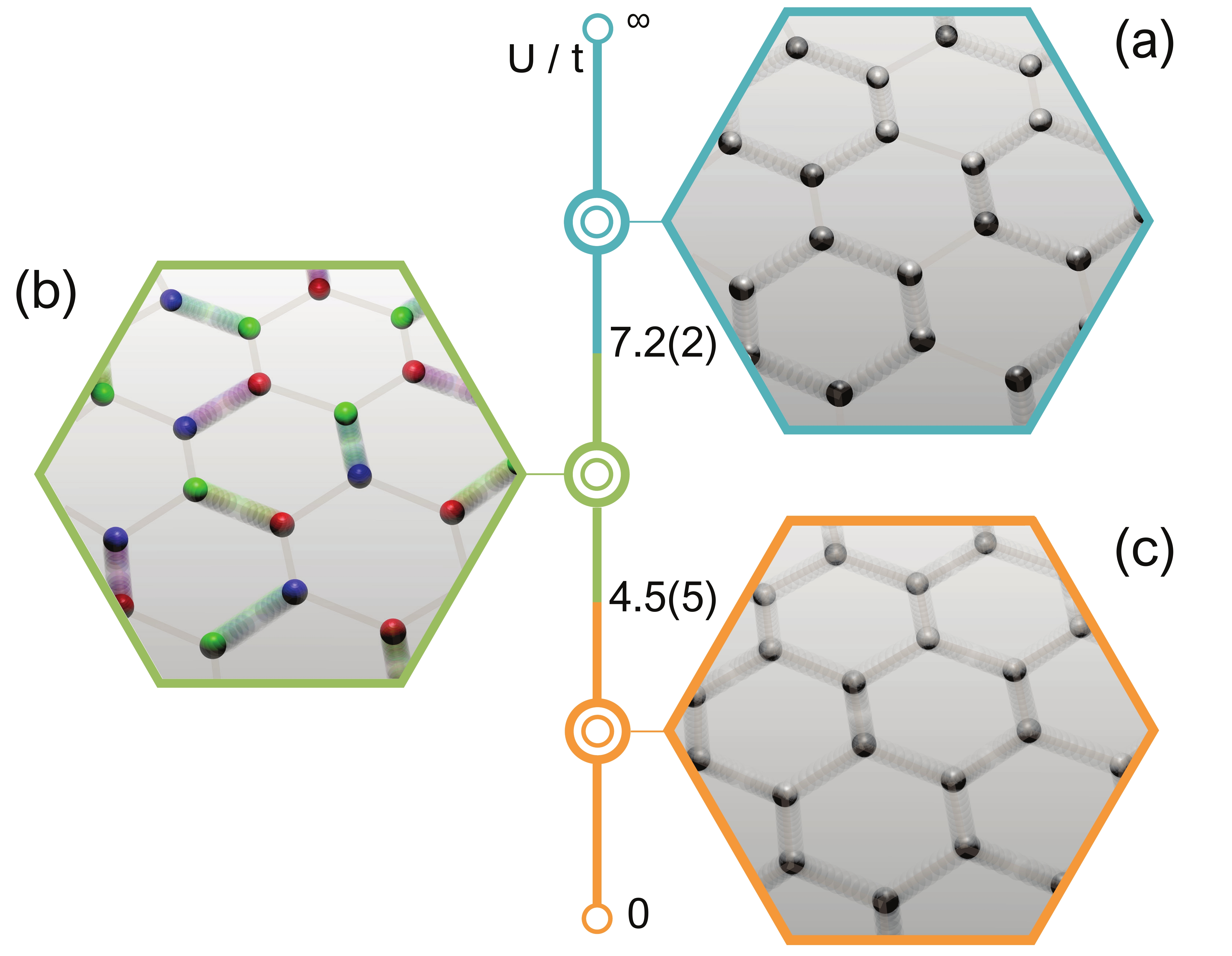}
 \caption{\label{fig:fig1} Ground state phase diagram of the SU(3) Hubbard model on the honeycomb lattice at 1/3 filling, including  (a) the plaquette phase, (b) the dimerized, color-ordered phase, and (c) the uniform (conducting) phase.} \end{figure}

In this paper we demonstrate that the 2D \sun Hubbard model has become within reach of state-of-the-art 2D tensor networks, which are a generalization of matrix product states to higher dimensions. Two-dimensional tensor networks have already been successfully applied in the \sun Heisenberg case~\cite{corboz11-su4,Bauer12,Zhao2012,Corboz12_simplex,Corboz12_su4,Corboz2013,nataf16} and the SU(2) Hubbard model~\cite{zheng17} (and many other strongly correlated systems; see, e.g., Refs.~\onlinecite{Corboz2014b,Niesen2017a,Liao2017,Haghshenas2018,chen18,Lee2018,Jahromi2018} and references therein). 
Here we consider the SU(3) Hubbard model on a honeycomb lattice and determine the nature of the Mott insulating phases as a function of the interaction strength, summarized in the phase diagram shown in Fig.~\ref{fig:fig1}. Our main findings include  a plaquette ground state in the strongly interacting limit, consistent with the result for the Heisenberg case~\cite{Zhao2012,Corboz2013}, whereas in the intermediate interaction range another state which has a dimer order with a coexisting color order gets stabilized by enhanced charge fluctuations.

\section{Model} 
The \sun Hubbard model is given by 
\begin{equation}
\hat{H}=-t \sum_{\left\langle i,j\right\rangle ,\alpha}\hat{c}_{i\alpha}^{\dagger}\hat{c}_{j\alpha}+H.c.+U\sum_{i,\alpha<\beta}\hat{n}_{i\alpha}\hat{n}_{i\beta},
\end{equation}
where $\alpha, \beta = 1...N$ label the $N$ different colors (or flavors) of fermions, $i$ and $j$ denote the lattice sites, and $\left\langle i,j\right\rangle$ runs over pairs of nearest neighbor sites. The operator $\hat{c}_{i\alpha}^\dagger$ ($\hat{c}_{i\alpha}$) creates (annihilates) a fermion of type $\alpha$ at site $i$, and $\hat{n}_{i\alpha}\equiv\hat{c}_{i\alpha}^\dagger\hat{c}_{i\alpha}$ is the particle number operator of type $\alpha$.  Here we consider the model for $N=3$ at 1/3 filling (one particle per site) on a honeycomb lattice and study its phase diagram as a function of $U/t$.

\section{Method}
An infinite projected entangled pair state (iPEPS) is a variational tensor network ansatz to efficiently represent 2D ground states of local Hamiltonians in the thermodynamic limit~\cite{Verstraete2004,
Jordan2008,nishio2004}.  It consists of a unit cell of tensors, with one tensor per lattice site, which is periodically repeated on the lattice. Each tensor has one physical index of dimension $d$ representing the local Hilbert space of a lattice site (here $d=8$), and $Z$ auxiliary indices with a bond dimension $D$ where $Z$ is the coordination number of the lattice ($Z=3$ for the honeycomb lattice), and $D$ controls the accuracy of the ansatz. 
 
For our simulations, we use the fermionic iPEPS ansatz from Ref.~\onlinecite{corboz2010} (see also Refs.~\onlinecite{kraus2010,Barthel2009,Corboz10_fmera,pineda2010,Corboz09_fmera,shi2009}) with unit cells up to $18$ sites. The corner transfer matrix method \cite{Baxter1976,Nishino1996,Orus2009,Corboz2011,corboz14_tJ} is used to contract the infinite 2D network of tensors, where we map the honeycomb lattice onto a brick-wall square lattice~\cite{Corboz12_su4}. The optimal variational parameters to approximate the ground state are obtained based on an imaginary time evolution~\cite{corboz2010,Phien2015}. The truncation of a bond index, required at each imaginary time step, is done using both the so-called \emph{simple-update} \cite{Jiang2008} and the more accurate (but computationally more expensive)  \emph{full-update} scheme~\cite{corboz2010,Phien2015}.  To increase the efficiency we use  tensors with  $U(1)\times U(1)\times U(1)$ symmetry, which significantly reduces the number of variational parameters, enabling us to reach large bond dimensions up to $D=28$. 


\section{Results}
\begin{figure} 
 \includegraphics[width=\columnwidth]{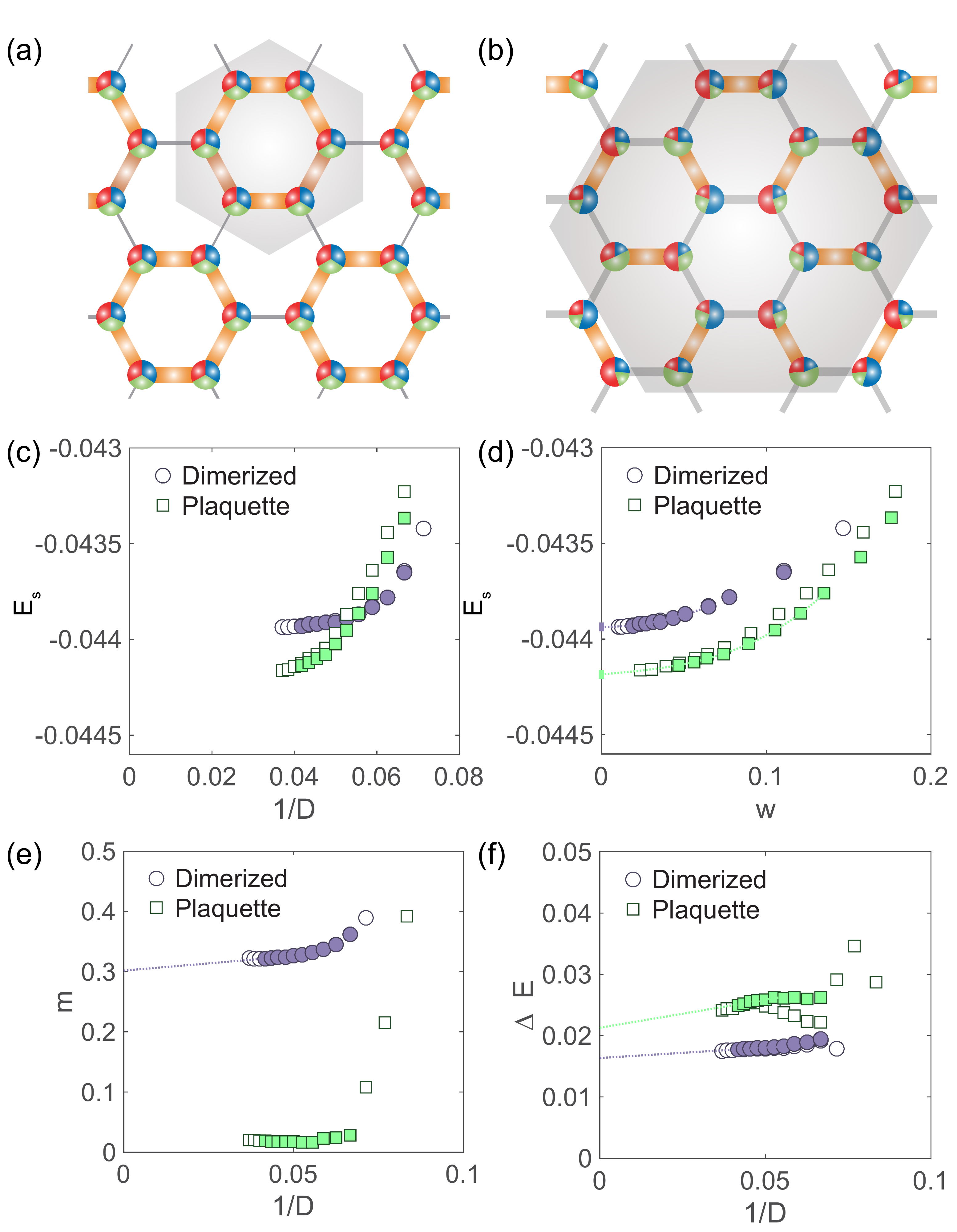}
 \caption{\label{fig:fig2} Graphical representations of the local expectation values for (a) the plaquette state and (b) the dimerized, color-ordered state, both at $U/t=100$, obtained from iPEPS simulations with a bond dimension of $D=25$.  Pie charts indicate the local color densities.  Thick (thin) bonds indicate low (high) bond energies; the strong bonds of the plaquettes and dimers are highlighted in orange for clarity.  The shaded hexagon in (a) encompasses the 6-site unit cell used to represent the plaquette state, whereas an 18-site unit cell is used in (b) to describe the dimerized state. (c, d) Average energies per site as functions of the inverse bond dimension (c) and the truncation error $w$ (d) for the two competing states at $U/t=100$.  Filled and open symbols correspond to full-update  and simple-update simulation results, respectively. (e)~Local ordered moment $m$ and (f) maximal difference in bond energies as a function of the inverse $D$. Dashed lines are guides for the eye.
 }   %
 \end{figure}

We first focus on the strongly interacting limit ($U/t=100$) where we find results which are consistent with those from the previously studied SU(3) Heisenberg model,~\cite{Zhao2012,Corboz2013} namely, a ground state with plaquette order in a six-site unit cell depicted in Fig.~\ref{fig:fig2}(a). The thickness of the bonds is proportional to the square magnitude of the corresponding local energies (called bond energies $E_b$), and the pie charts show the local densities of each color, which sum to $n=1$. In this case, all color densities are equal (i.e. $n_\alpha=1/3$) showing the absence of color order [i.e. SU(3) symmetry is not broken], and strong energy bonds are found around plaquettes consistent with  SU(3) singlets which are predominantly formed around the hexagons and thereby break translational symmetry. 
As in the Heisenberg case~\cite{Lee2012,Corboz2013} we find a competing low-energy state in a 18-site unit cell which is dimerized and has color order, shown in Fig.~\ref{fig:fig2}(b). Each dimer (in orange)  has two colors that are dominant at each end, and at the four sites surrounding a dimer the remaining third color is dominant. 
%

In Fig.~\ref{fig:fig2}(c) we compare the energies of the two states as a function of $1/D$ using both simple- and full-update optimizations. For large bond dimensions we clearly find that the plaquette state is energetically  favorable. The same conclusion is reached by plotting the energy as a function of the truncation error $w$, shown in Fig.~\ref{fig:fig2}(d), which often provides a better parameter for an extrapolation of the energy to the exact infinite-$D$ limit~\cite{Corboz2016}. The extrapolated energy of the dimer state, $E_s=-0.043936(9)$, is clearly higher than that of the plaquette state, $E_s=-0.044184(9)$.

In Fig.~\ref{fig:fig2}(e) we present  results for the local ordered moment $m$, given by 
\begin{equation}
m_{i}=\sqrt{\frac{3}{2}\sum_{\alpha=1}^{3}\left(\left\langle n_{i,\alpha}\right\rangle -\frac{1}{3}\right)^{2}}, 
\end{equation}
averaged over all lattice sites $i$ in the unit cell, which is finite in the dimer state and vanishing in the plaquette state. Figure~\ref{fig:fig2}(f) shows the difference in energy between the highest and the lowest bond energy in the unit cell, $\Delta E = \max(E_b) - \min(E_b)$, which is finite for both states. Since the large-$D$ results between simple and full updates are found to be similar for both states, we continue using the former, since it allows us to reach larger $D$ values with substantially smaller computational effort.

\begin{figure} 
\includegraphics[width=\columnwidth]{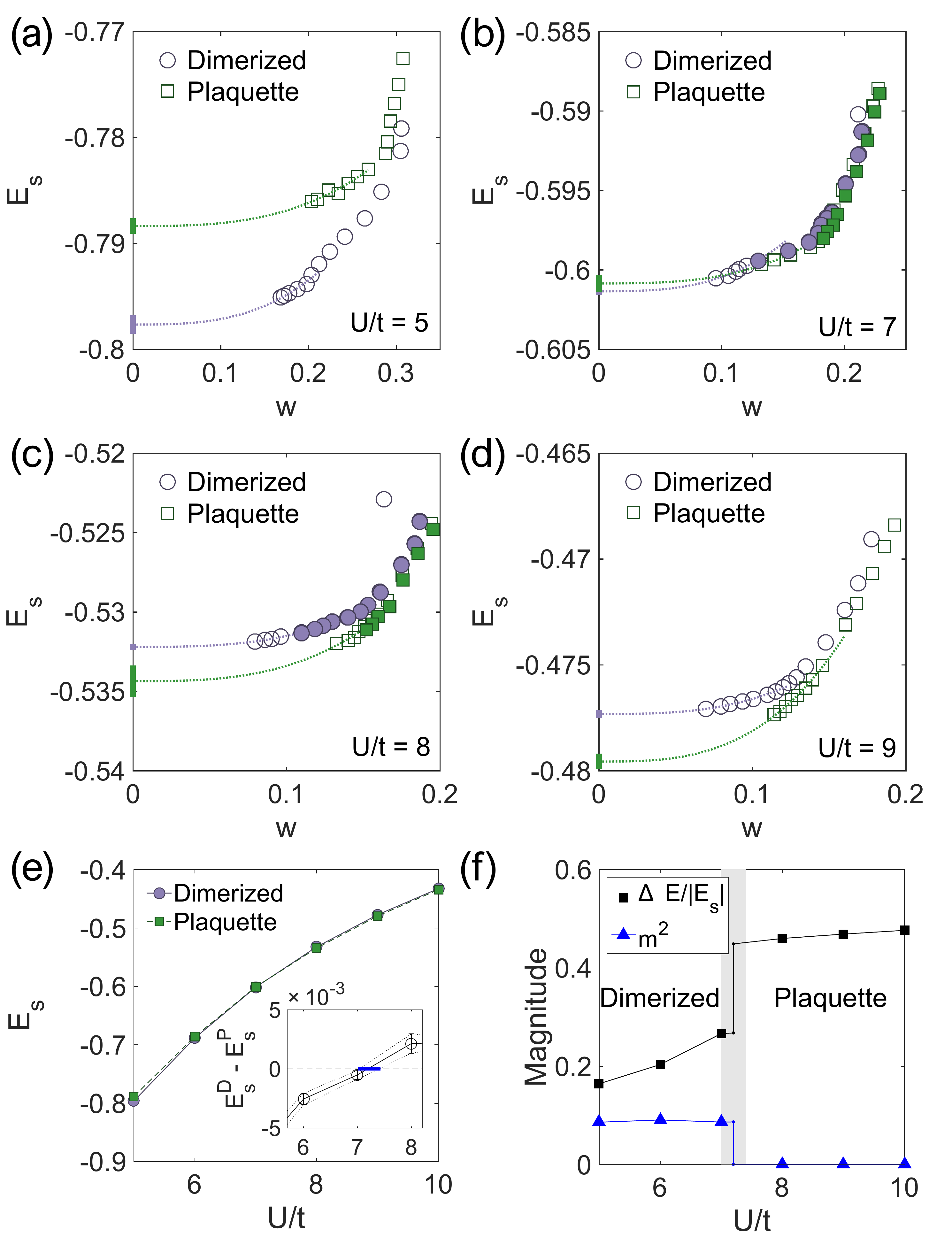}
\caption{\label{fig:fig3} 
Energy per site as a function of the truncation error of the two competing states for different values of $U/t$ (simple update). (a) For $U/t=5$ the dimerized state is clearly lower in energy, whereas (c) for $U/t=8$  and (d) for $U/t=9$ the plaquette state is lower. (b) For $U/t=7$ the extrapolated energies are very close. As a comparison we also show full-update results (filled symbols) yielding similar energies. (e) Extrapolated energy per site of the two states as a function of $U/t$. Inset:  difference in energy per site of the two states. By linear interpolation, taking into account the extrapolation errors, we find a critical value $U_c/t=7.2(2)$. (f) Order parameters of the ground state as a function of $U/t$ (for $D=27$) exhibiting a jump at the critical point. The shaded region marks the uncertainty of the location of the critical point.
}
\end{figure}

We next move away from the Heisenberg limit by reducing $U/t$ which enhances charge fluctuations. Interestingly, we observe that the energy difference between the two competing states gets smaller with decreasing $U/t$ [see Figs.~\ref{fig:fig3}(a)-\ref{fig:fig3}(d), and around a critical value of $U_c/t = 7.2(2)~$\footnote{This value, which is based on an extrapolation in the truncation error $w$, is compatible with the one obtained using a $1/D$ extrapolation, $U/t=7.4(7)$.}  the energies of the two states intersect [Fig.~\ref{fig:fig3}(e)] so that the dimer state actually becomes the ground state for $U<U_c$~\footnote{We also tested other unit cell sizes but could not find other competing states, except deformed versions of the color-ordered, dimerized state which has a higher energy}. Consequently, at the critical point, both $m$ and $\Delta E$  exhibit a discontinuity [Fig.~\ref{fig:fig3}(f)], characteristic for the first-order nature of the transition.

 Thus, enhanced charge fluctuations present at a lower $U/t$ are found to favor the dimerized state over the plaquette state. Intuitively this is because in the latter the kinetic energy is mostly concentrated locally around the plaquettes, whereas in the dimer state charges are delocalized more uniformly in the system [which can be seen from the smaller value of $\Delta E$ in the dimerized state than in the plaquette state; see Fig.~\ref{fig:fig2}(f)], which is favored as we decrease the interaction (with the fermions being completely delocalized in the non-interacting limit).

\begin{figure}[tb]
\includegraphics[width=\columnwidth]{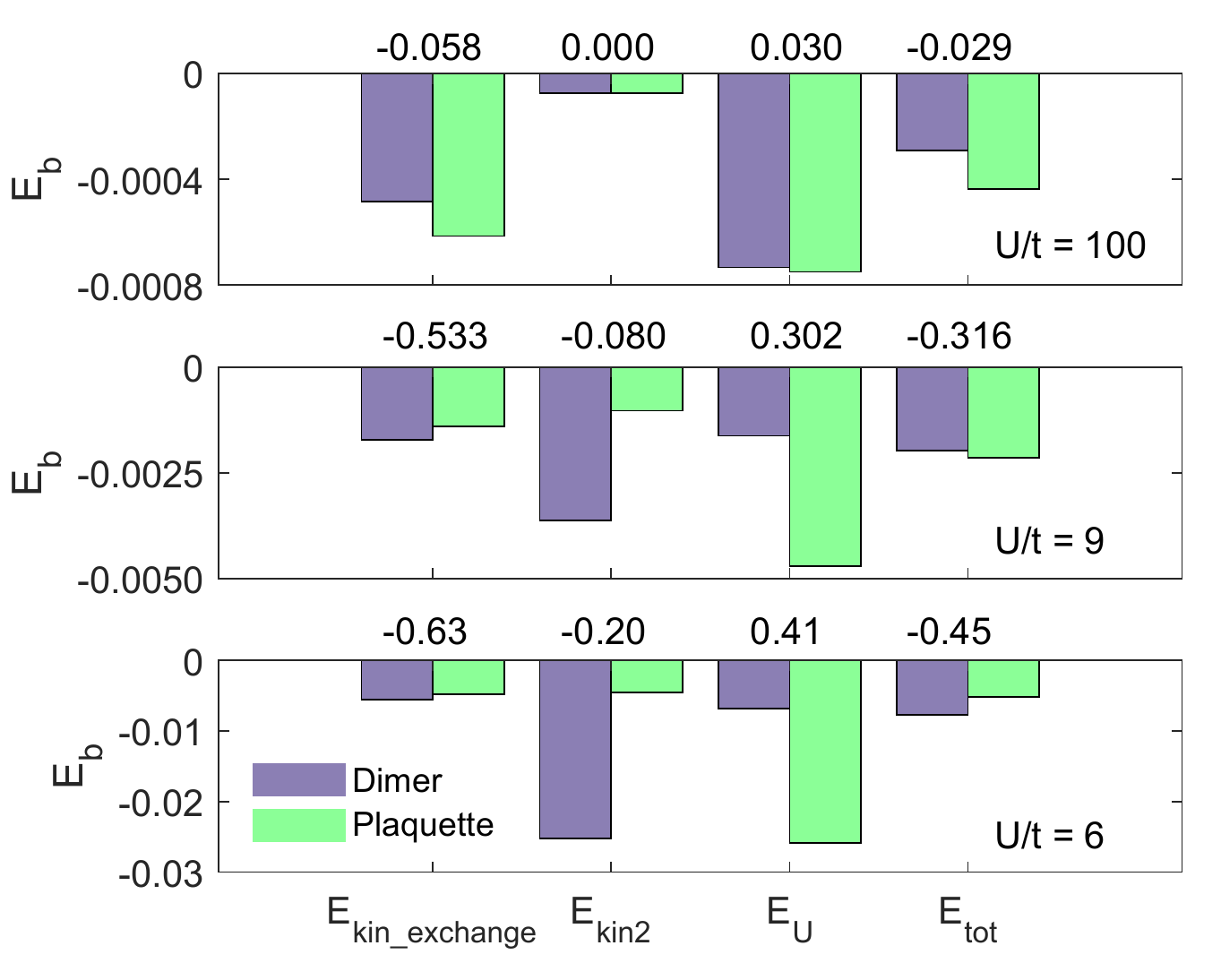}
\caption{\label{fig:Econtr} 
Comparison of the energy contributions per bond of the two competing states for different values of $U/t$ ($D=26$, simple update). The total kinetic energy is split into a term relevant for the Heisenberg super-exchange and a second term containing the remaining contributions (see text). All energies have been shifted by the values shown at the top of each pair of bars.}
\end{figure}

To get more insights into the energetics of the two states we present a comparison of the different energy contributions in Fig.~\ref{fig:Econtr}  for $U/t=100$, $U/t=9$, and $U/t=6$. The kinetic  term is split into two parts. The first part is the one relevant for the superexchange processes  in the Heisenberg (large-$U/t$) limit, i.e., matrix elements between two singly occupied sites $|1_\alpha,1_\beta \rangle$ and an empty and doubly occupied site  $|0,1_\alpha 1_\beta \rangle$ (or $|1_\alpha 1_\beta ,0\rangle$), where $\alpha \neq \beta$ denote the colors of the two fermions. The second term $E_{kin2}$ includes all remaining kinetic contributions. In the Heisenberg limit the latter can be neglected (see top panel in Fig.~\ref{fig:Econtr}), and the plaquette state wins over the dimerized state thanks to a slightly lower exchange energy. When $U/t$ is lowered the energy cost to form doubly occupied sites decreases, and enhanced charge fluctuations lead to a gain in kinetic energy, including also contributions which are not included in the exchange term. These contributions are clearly stronger in the dimer state, at the expense of a larger positive contribution from the on-site repulsion (see middle panel in Fig.~\ref{Fig:Econtr}). Eventually, for even smaller $U/t$ these kinetic contributions become dominant compared to the on-site repulsion, and the dimerized state wins over the plaquette state (bottom panel in Fig.~\ref{Fig:Econtr}). 

When $U/t$ is lowered even further one can expect that the system will eventually undergo a transition from the Mott insulating dimer state into a conducting (or superconducting) state at weak~$U/t$. In the non-interacting case, $U/t=0$, the model reduces to a tight-binding model of three independent species of fermions. In contrast to the half-filled case where the Fermi level crosses the Dirac nodes (as, e.g., in graphene), the Fermi surface at 1/3 filling is one-dimensional manifold in  momentum space, which is known to lead to a multiplicative logarithmic correction to the area law of the entanglement entropy~\cite{wolf2006,gioev06}. Thus, in the low-$U/t$ limit the states are strongly entangled  and therefore very challenging to  accurately  represent by a tensor network ansatz. This is also reflected in the increasing truncation error with decreasing $U/t$ (for a fixed value of $D$; see Fig.~\ref{fig:fig3}).

\begin{figure}[tb]
\includegraphics[width=\columnwidth]{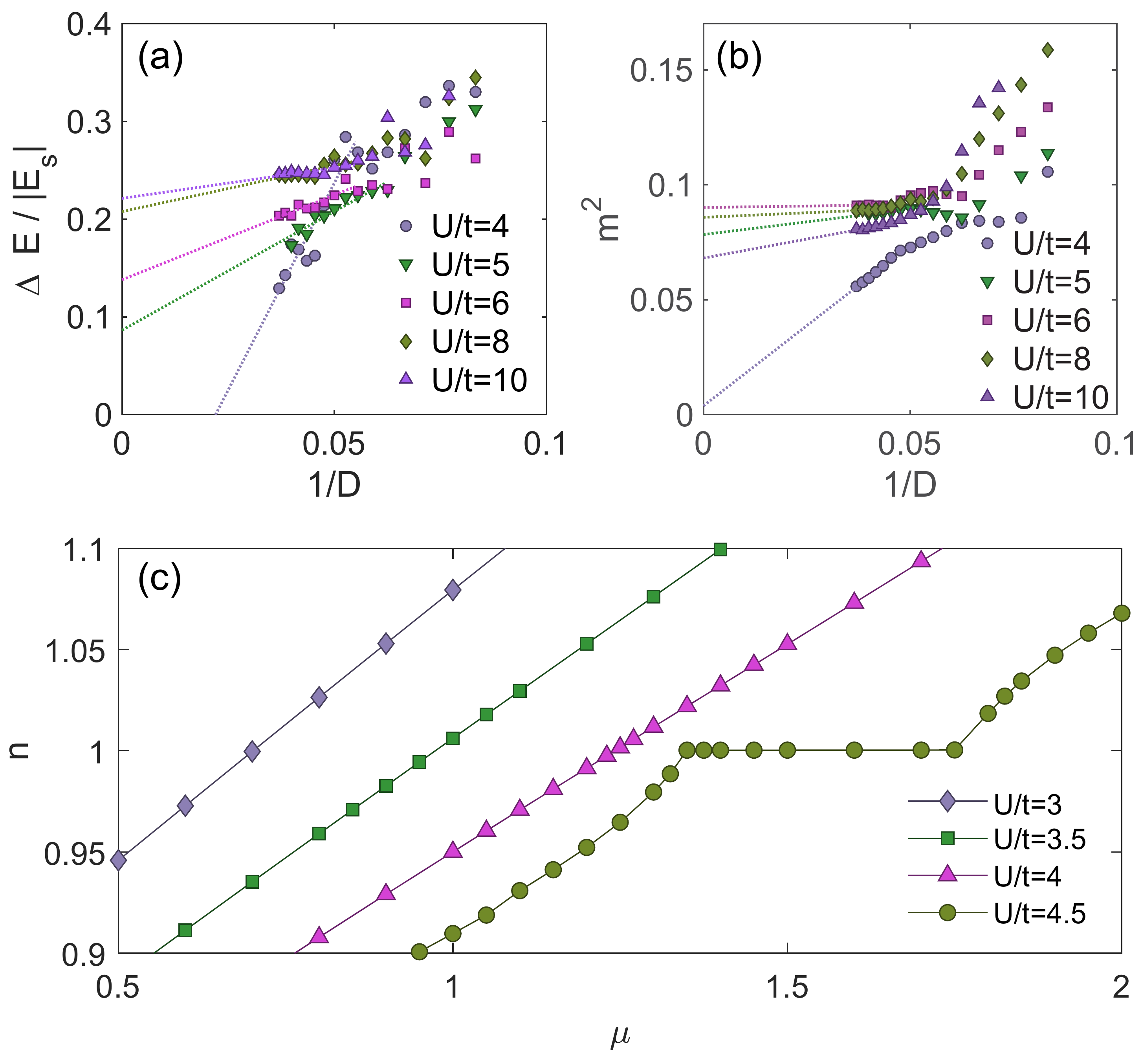}
 \caption{(a) Relative energy difference between the maximum and the minimum bond energies and (b) the square of the local ordered moment as a function of the inverse $D$ of the dimerized state. For $U/t=4$ the strong suppression of the dimer and color order suggest that the dimerized state becomes unstable (i.e., that the true ground state is uniform). (c) Particle density as a function of the chemical potential of a uniform state obtained with a two-site iPEPS ansatz ($D=28$; simple update). The charge gap clearly vanishes for $U/t\le4$. \label{fig5} }
\end{figure}

To obtain an estimate of the Mott transition we study the stability of the dimer phase upon lowering $U/t$. The  energy difference between the highest and the lowest bond energies shown in Fig.~\ref{fig5}(a) gets strongly suppressed with increasing $D$ for $U/t \le 5$. Taking a linear extrapolation in $1/D$ as a rough estimate for the value in the infinite-$D$ limit, we find that $\Delta E$ --- where $\Delta E > 0$ indicates breaking of translational symmetry --- vanishes for $U/t=4$. Also the square of the local ordered moment --- where $m^2>0$ indicates breaking of SU(3) color symmetry --- shown in Fig.~\ref{fig5}(b) extrapolates to a vanishingly small value in the infinite-$D$ limit for $U/t=4$.  From these results we conclude that the dimer state gets unstable between $U/t=4$ and $U/t=5$, i.e. that the ground state becomes uniform for $U/t<U_c/t=4.5(5)$. Furthermore, we also find that the charge gap of the uniform state, obtained using a two-site unit cell iPEPS ansatz, clearly vanishes for $U/t=4$ [see Fig.~\ref{fig5}(c)]. For $U/t=4.5$ and $D=28$ the charge gap is still visible by the plateau in $n(\mu)$, but its size gets suppressed with increasing $D$. This suggests that the Mott transition also takes place at $U_c$ (i.e.,  there is no clear indication of an additional intermediate quantum spin liquid insulating phase).~\footnote{We note that determining the nature of the transition point between the dimer and uniform state is challenging. Due to the increased entanglement with decreasing $U/t$ it is difficult to distinguish a continuous transition from a weak first-order transition here}. 

%

\section{Conclusions}
Using iPEPS simulations pushed to large bond dimensions, up to $D=28$, we have investigated the ground-state phase diagram of the SU(3) Hubbard model on a honeycomb lattice at 1/3 filling, focusing on the Mott insulating phases. 
In the large-$U/t$ (Heisenberg) limit we found a plaquette ground state which breaks lattice symmetry but preserves \sun symmetry. For an intermediate interaction strength, $4.5(5)\le U/t \le 7.2(2)$, the ground state is a dimerized, color-ordered state which breaks both \sun and lattice symmetry. The order parameters of the dimerized state vanish for $U \le 4.5(5)$, indicating that the ground state becomes uniform, compatible with a conducting (or superconducting) state in the weakly interacting limit. 

This work demonstrates that the 2D \sun Hubbard model in the strongly correlated regime has become within reach of state-of-the-art tensor network methods, offering a systematic and controlled way to predict the phases of these challenging models. By making use of global symmetries it is also  possible to study systems with a larger $N$, despite the  large local dimension of $d=2^N$.
 
Our results may serve as a prediction for future quantum simulators based on SU(3) ultra-cold atoms in an optical honeycomb lattice. Since the plaquette phase breaks a discrete lattice symmetry we can expect that it extends also to finite temperatures, so that the phase can be realized in experiments at sufficiently low temperatures,  offering also an interesting possibility to study a finite-temperature phase transition in a 2D quantum simulator. While the finite color order of the dimer phase can only exist at zero temperature in two dimensions, the dimerization may in principle set in already at finite temperature (without coexisting color order), or occur simultaneously with the color order at zero temperature. Recently developed tensor network approaches for finite-temperature simulations~\cite{czarnik12,czarnik14,czarnik15b,czarnik16b,kshetrimayum18,czarnik19} may provide further insights into the critical temperatures in the future.

\begin{acknowledgments}
S.S.C. acknowledges helpful discussions with I.~Niesen. This project received funding from the European Research Council (ERC) under the European Union's Horizon 2020  research and innovation program (Grant Agreement No.~677061). This work is part of the \mbox{Delta-ITP} consortium, a program of the Netherlands Organization for Scientific Research (NWO) that is funded by the Dutch Ministry of Education, Culture and Science~(OCW).
\end{acknowledgments}

\bibliographystyle{apsrev4-1}
\bibliography{refs,biblio}
\end{document}